\documentclass[journal]{vgtc}                     

\onlineid{0}

\vgtccategory{Research}

\vgtcpapertype{evaluation}

\title{A Multidimensional Assessment Method for Situated Visualization Understanding (MdamV)}

\author{
  \authororcid{Antonia Saske}{0009-0005-4250-8464}, \authororcid{Laura Koesten}{0000-0003-4110-1759},  \authororcid{Torsten Möller}{0000-0003-1192-0710}, \authororcid{Judith Staudner}{0009-0002-6636-7975}, and \authororcid{Sylvia Kritzinger}{0000-0003-2765-8200}}

\authorfooter{
   \item
    Antonia Saske, 
    	E-mail: antonia.saske@univie.ac.at.
  \item 
    Laura Koesten, 
  	E-mail: laura.koesten@univie.ac.at.
 \item 
    Torsten Möller,
        E-mail: torsten.moeller@univie.ac.at.
    
  \item 
    Judith Staudner, 
  	E-mail: judith.staudner@univie.ac.at.
  \item 
    Sylvia Kritzinger, 
  	E-mail: sylvia.kritzinger@univie.ac.at.}

\abstract{
How audiences read, interpret, and critique data visualizations is mainly assessed through performance tests featuring tasks like value retrieval. Yet, other factors shown to shape visualization understanding, such as numeracy, graph familiarity, and aesthetic perception, remain underrepresented in existing instruments. To address this, we design and test a Multidimensional Assessment Method of Situated Visualization Understanding (MdamV). This method integrates task-based measures with self-perceived ability ratings and open-ended critique, applied directly to the visualizations being read. Grounded in learning sciences frameworks that view understanding as a multifaceted process, MdamV spans six dimensions: \textit{Comprehending}, \textit{Decoding}, \textit{Aestheticizing}, \textit{Critiquing}, \textit{Reading}, and \textit{Contextualizing}. Validation was supported by a survey (N=438) representative of Austria's population (ages  18-74, male/female split), using a line chart and a bar chart on climate data. Findings show, for example, that about a quarter of respondents indicate deficits in comprehending simple data units, roughly one in five people felt unfamiliar with each chart type, and self-assessed numeracy was significantly related to data reading performance (p=$0.0004$). Overall, the evaluation of MdamV demonstrates the value of assessing visualization understanding beyond performance, framing it as a situated process tied to particular visualizations.

}


\graphicspath{{figs/}{figures/}{pictures/}{images/}{./}} 

\usepackage{tabu}                      
\usepackage{amsmath}
\usepackage{makecell}
\usepackage{mathptmx}

\begin{document}



\maketitle

\section{Motivation}

When interpreting data visualizations, audiences draw on a range of related abilities to read, interpret, and critically engage with them~\cite{kim2017, brill2007}. This range can be referred to as visual data literacy and has typically been assessed through performance tests. Well-known examples include the Visualization Literacy Assessment Test (VLAT)~\cite{kim2017}, its briefer version, the Mini-VLAT~\cite{pandey2023}, or the Critical Thinking Assessment for Literacy in Visualizations (CALVI)~\cite{ge2023}, all of which focus on evaluating correct responses to tasks, such as value retrieval. While these instruments are useful for treating literacy as a measurable ability, they fail to capture visualization understanding beyond a strong focus on quantifying test takers' accuracy in selected-response items. Research has established that facets like graph familiarity~\cite{fischer2018} and topic knowledge~\cite{boykoff2011belief} can shape reading accuracy, and shows that literacy is often approached as a multi-faceted set of abilities \cite{kucer2009multidimensional, kucer2015literacy, panos2021multidimensional}. Still, such insights remain underutilized in ability assessments.

To address this gap, we frame visualization understanding as the application of visual data literacy in context-specific settings, meaning, in direct relation to the visualizations used in the study. We present a Multidimensional Assessment Method of Situated Visualization Understanding (MdamV) that was applied to static climate data visualizations. MdamV was developed in alignment with learning sciences theory on understanding as a multifaceted process. It was then validated through qualitative pilots with eight participants and a survey (N=$438$) representative of the Austrian population's age groups from 18 to 74 years and their male/female gender split. Our method combines task-based measures with self-perceived ability ratings and open-ended critique. This combination captures not only whether respondents can complete tasks correctly, but also how they perceive their own understanding and indicate difficulties with specific visualization features. Therefore, it reveals nuances that established assessments may miss, such as perceived trustworthiness or challenges with decoding color. MdamV is composed of six assessment dimensions:

\begin{itemize}[noitemsep]
     \item \textit{Comprehending}: Overall Impression and Abstract Thinking,
     \item \textit{Decoding}: Graph Component Understanding,
     \item \textit{Aestheticizing}: Aesthetic Perception,
     \item \textit{Critiquing}: Visualization Critique,
     \item \textit{Reading}: Low-level Data Reading Tasks,
     \item \textit{Contextualizing}: Numeracy and Topic Knowledge.
 \end{itemize}

Designed for general audiences, MdamV addresses some limitations of prior assessments that strongly focused on expert audiences~\cite{fischer2018, harold2019} or specific groups, such as students~\cite{callow2008, firat2023constructivism}. Given the increasing need for data literacy---and not everyone being equally data-savvy~\cite{lee2020}---our survey targeted a broad audience, including people of varying ages and without expertise in data visualization.

Here, we showcase the design of MdamV and report on its application in a survey representative of Austria's population (ages 18-74, male/female split). Our results demonstrate that a multidimensional approach can yield informative insights into audience understanding beyond visualization task performance. They also show how respondents engaged with two tested visualizations, a line chart and a bar chart on climate data, chosen for their simplicity, familiarity, and importance in conveying critical information~\cite{moser2010communicating, o2014climate, schuster2022}. As MdamV is evaluated with two specific visualizations, it should not be seen as a ready-to-use instrument but rather a method with the potential to be adapted and applied in different contexts.
 
In our representative survey, we found that $48\%$ of respondents in Austria make mistakes with the simple line and bar chart. About a quarter of respondents found simple data units, e.g., millimeters, in the surveyed charts incomprehensible. Further, about $12\%$ felt unfamiliar with both displayed chart types. Concerning aesthetic perception, there was a tendency to find the visualization pretty or well-designed and simultaneously rate it as trustworthy. In optional text fields, $37\%$ of survey respondents reported at least one point of critique, offering qualitative insights into both their critical engagement and the visualization elements they found irritating. In our survey, respondents' self-assessed numeracy level relates significantly (p=$0.004$) to data reading performance.

\begin{table*}[ht]
\centering
\renewcommand{\arraystretch}{2}
\begin{tabular}{l l l l l}
\makecell[hl]{\textbf{Assessment Dimension} \\ \textbf{in MdamV}} & \makecell[hl]{\textbf{Perspective}} & \makecell[hl]{\textbf{Ability Examples Related to} \\ \textbf{Visual Data Literacy}} & \makecell[hl]{\textbf{Related Facet of} \\ \textbf{Understanding~\cite{wiggins2005}}} & \makecell[hl]{\textbf{Exemplary Findings From} \\ \textbf{Our Representative Survey}}  \\ \hline
Dim1: Comprehending & Top-Down & \makecell[hl]{Summarize visualized content, recall\\ information~\cite{bowen2017, harold2019}} & Explanation & \makecell[hl]{$76\%$ feel confident in summa-\\rizing} \\ \hline
Dim2: Decoding & Bottom-Up & \makecell[hl]{Examine color usage, analyze graph\\ elements~\cite{robinson1984, fischer2018}} & Interpretation & \makecell[hl]{$12\%$ feel unfamiliar with both\\ line and bar charts}\\ \hline
Dim3: Aestheticizing & Bottom-Up & \makecell[hl]{Decide upon trustworthiness, judge\\visual appeal~\cite{chujun2021, he2023}} & Empathy & \makecell[hl]{Perceived beauty was linked to \\ trust}\\ \hline
Dim4: Critiquing & Bottom-Up & \makecell[hl]{Articulate critical thought, recognize\\the meaning of construction\cite{brill2007, chevalier2018}} & Perspective & \makecell[hl]{$37\%$ submitted meaningful\\critique in optional inputs} \\ \hline
Dim5: Reading & Top-Down & \makecell[hl]{Solve data reading tasks~\cite{kim2017, pandey2023}} & Interpretation & \makecell[hl]{$48\%$ made errors in reading\\ tasks} \\ \hline
Dim6: Contextualizing & Top-Down & \makecell[hl]{Utilize statistical skill, put visual-\\ization in context~\cite{xexakis2021, boykoff2011belief}} & \makecell[hl]{Application} & \makecell[hl]{Self-assessed numeracy corre-\\lated with data reading} \\
\end{tabular}
\caption{Assessment dimensions covered in MdamV, developed from either a viewer-focused (top-down) or visualization-focused (bottom-up) perspective. Each dimension is illustrated by example activities and aligned with a\textit{ Facet of Understanding}~\cite{wiggins2005}. Additionally, we share insights from our representative survey where we applied MdamV to two climate data visualizations. The results are analyzed in detail in Section~\ref{sec:results}.}
\label{tab:dimensions-and-wiggins}
\end{table*}

\section{Background}
Through a theory-driven approach, we developed six different assessment dimensions for visualization understanding, drawing on prior work related to visual data literacy. We frame visualization understanding as the situated application of visual data literacy: a process in which a range of abilities come together to understand a particular visualization. Therefore, we consider both the reader's perspective and the characteristics of the visualization~\cite{hegarty2005, robinson1984, michelson2017}. Building on this, we align our assessment dimensions with learning sciences frameworks that define \textit{understanding} as "a family of related abilities"~\cite{wiggins2005}. Creating a multidimensional approach was deemed necessary because current assessments are dominantly task-based, and therefore, they are perspectively limited in their ability display. We highlight studies using self-perceived abilities in relation to performance, arguing that such ratings can offer unique insights into visualization understanding beyond established tests. 

\subsection{Visual Data Literacy: Viewer and Visualization Perspective}
\label{sec:perspective}

Visual literacy has been explored as an intellectual and social phenomenon since the 1960s, with early contributions emerging in fields such as education, communication, visual arts, and psychology~\cite{michelson2017}, addressing both audience-related factors and properties of visual design. Accordingly, it can be viewed through a bipartite perspective, distinguishing between top-down and bottom-up aspects~\cite{hegarty2005}. Top-down characteristics take the perspective of visualization readers and are shaped by factors such as prior beliefs, demographic characteristics, and personal skills. Here, numeracy, for instance, has been identified as a critical skill influencing data reading accuracy and knowledge gains~\cite{xexakis2021}. In contrast, bottom-up characteristics focus on the visualization itself, including its content or aesthetic structure~\cite{robinson1984}. We argue that it is valuable to capture visualization understanding through dimensions that consider the influence of visualization design on audience responses. Visualization techniques, such as color choices or the depiction of uncertainty, have been shown to directly influence how audiences retrieve information and make decisions~\cite{munzner_visualization_2014, thompson_influence_2015, fang_evaluating_2022}.

Since we recognize both top-down and bottom-up aspects, we align our assessment dimensions with either a viewer- or visualization-centered focus, as shown in \autoref{tab:dimensions-and-wiggins}. For example, questions on \textit{Contextualizing} represent a top-down perspective and include viewers' numeracy and topic interest, while questions on \textit{Decoding} reflect a bottom-up perspective by referring to color encoding and chart type. 

\subsection{Deconstructing Visualization Understanding into Ability Facets}
\label{sec:deconstruct}

Insights from learning sciences, such as Wiggins and McTighe's framework of the \emph{Six Facets of Understanding}~\cite{wiggins2005}, conceptualize \emph{understanding} as a multifaceted process composed of different abilities: explaining, interpreting, applying, empathizing, having perspective, and self-knowledge. These facets can be aligned with representing different abilities that are applied during visualization understanding. For example, a respondent who can summarize the content of a data visualization (see Q1c in Fig. \ref{fig:overview-batteries-questions}) would be able to explain. If a respondent solves a data reading task (Q5a-d) or translates color encoding (Q2b), this is an act of interpretation. Deciding upon the trustworthiness of a visualization (Q3d) requires empathizing with its creator's intent. Critiquing or suggesting design changes (Q4b/c) indicates perspective-taking. Someone who can contextualize displayed data (Q6b) can apply, e.g., by drawing on their knowledge of the visualization's topic. Finally, self-knowledge involves consciously questioning one's own abilities and limits in understanding. While this could be valuable to assess, it lay beyond the scope of our survey design, e.g., by considering readers’ intrinsic habits of mind and how these shaped their engagement.


Building on these categories from the learning sciences, we aligned assessment dimensions with facets of \textit{understanding} to illustrate our procedure and to better exemplify visualization understanding as a process involving a multidimensional set of abilities (see Tab.~\ref{tab:dimensions-and-wiggins}).

Another learning sciences framework worth considering is \emph{Bloom's Taxonomy of Educational Objectives for Knowledge-Based Goals}~\cite{bloom1956}, which deconstructs learning into ability levels (memorizing, understanding, application, analysis, evaluation, and creation), which are most commonly used for cognitive domains~\cite{lee2023affectivelearning}. While our alignment focused on the \textit{Six Facets of Understanding}, this taxonomy offers a complementary view, for example, by linking the recall of information (Q1c) to memorizing, the salvation of a data reading task (Q5a-d) to application, or the examination of color encoding (Q2b) to analysis. It has already been adapted to assess literacy in contexts such as parallel coordinates plots by Peng et al.~\cite{peng2022evaluating}.

\subsection{Limitations of Prior Assessments}
\label{sec:background-limitations}
Prior assessments related to visual data literacy often rely on task-based approaches, focus on a single data visualization type, or target specific audiences. Examples include treemap and parallel coordinate plot literacy tests~\cite{firat2023constructivism, firat2020, elif2022}. Several studies addressed educational contexts, such as qualitative principles for assessing visual literacy in schools~\cite{callow2008}, while others examine expert understanding of climate data visualizations~\cite{fischer2018}. Despite these efforts, the utilized measurement approaches remain largely one-dimensional. Moreover, as noted in prior critiques, general audiences with varying ability levels are still underrepresented in this line of research~\cite{rogers2017, boerner2016, boerner2019}.

In 2017, Lee et al.~\cite{kim2017} introduced a Visualization Literacy Assessment Test (VLAT), a $53$-item multiple-choice test covering $12$ data visualizations. It specializes in measuring performance for simple visualization tasks like value retrieval and has been implemented or adapted multiple times~\cite{nobre2024vislit, hong2025vislit}. While effective for measuring task-based performance, VLAT's reliance on correct/incorrect responses limits its ability to capture other facets of visualization understanding, such as where deficits may lay and what visualization readers perceive as challenging. VLAT was recently condensed into the Mini-VLAT~\cite{pandey2023} to improve accessibility by reducing its original $20$+ minute duration. While this may improve accessibility, it still emphasizes numerical accuracy over other abilities. The Critical Thinking Assessment for Literacy in Visualizations (CALVI)~\cite{ge2023} builds on this approach by incorporating critical ability as part of visualization literacy. Comprising 45 items across nine chart types with intentional design flaws, CALVI meaningfully expands the assessment approach but remains focused on task performance. In other approaches, respondents are asked to construct visualizations, such as in the AVEC assessment~\cite{ge2025avec} or studies on parallel coordinate plot literacy~\cite{peng2022evaluating}, yet they too are primarily based on task performance.

\subsection{Self-Perception in Ability Assessments}
\label{sec:saps}

Recently, PREVis, a scale for perceived readability in data visualizations, was introduced by Cabouat et al.~\cite{cabouat2024previs} and exemplifies how self-assessment may provide insights into visualization understanding. Using Likert scales, it measures four facets of perceived readability, including visualization layout, and data value readability. PREVis provides a valuable entry point for respondents to self-assess the aspect of perceived readability~\cite{cabouat2024previs}. MdamV overlaps with some PREVis aspects, such as perceived understanding (see Q1a in Fig. \ref{fig:overview-batteries-questions}) and identifying visual elements (Q2a), but provides additional insights. For instance, our survey, detailed in Section \ref{sec:results}, found that $22\%$ of respondents indicated comprehension issues with millimeters, and $28\%$ with millions of tons (Q2a). This level of detail, tied to tested visualizations and their features, highlights comprehension challenges that PREVis would not capture. 

Self-assessment assumes that human traits and skills are quantifiable and measurable~\cite{hamilton2007} and can complement task-based tests to support more nuanced approaches to measuring abilities~\cite{allen2005, garcia2016}. In our method, self-assessment measures capture how visualization readers perceive facets of their visualization understanding. We hypothesize that these ratings can reveal nuanced insights into the respondents' set of abilities and may correlate with performance in data reading tasks. Similar approaches have been widely applied in psychology, education, and health research: self-concepts of talent have been compared to performance~\cite{bauer2023selfconcept}, self-perceptions have been linked to academic achievement~\cite{spinath2006selfperceived, greven2009spas}, perceived coping abilities have been shown to predict smoking cessation outcomes~\cite{nohlert2018selfperceived}. Further, self-perception is recognized as a survey approach in political science, i.e., to gain insights into respondents' stances and beliefs~\cite{morisi2019political, kritzinger2021trust}.

While self-perceived ability ratings differ from performance metrics and are not as "objective" or precisely quantifiable, they can still be predictive~\cite{alturki2022selfassessment}. Rather than replacing task-based instruments like VLAT~\cite{kim2017}, we integrate self-perceived ability ratings to broaden how visual data literacy is being approached. Beyond measuring what people \textit{can do}, self-assessment sheds light on how respondents \textit{perceive} their interactions with visualizations. In literacy assessments, factors related to the latter, such as how personal beliefs, trust, or emotional responses influence information retrieval, remain underexplored~\cite{davis2020pandemics, garfin2020novel}.

\section{MdamV: A Multidimensional Assessment Method for Situated Visualization Understanding}
\label{sec:Question Batteries}

\begin{figure}[ht]
    \centering
    \includegraphics[width=\columnwidth]{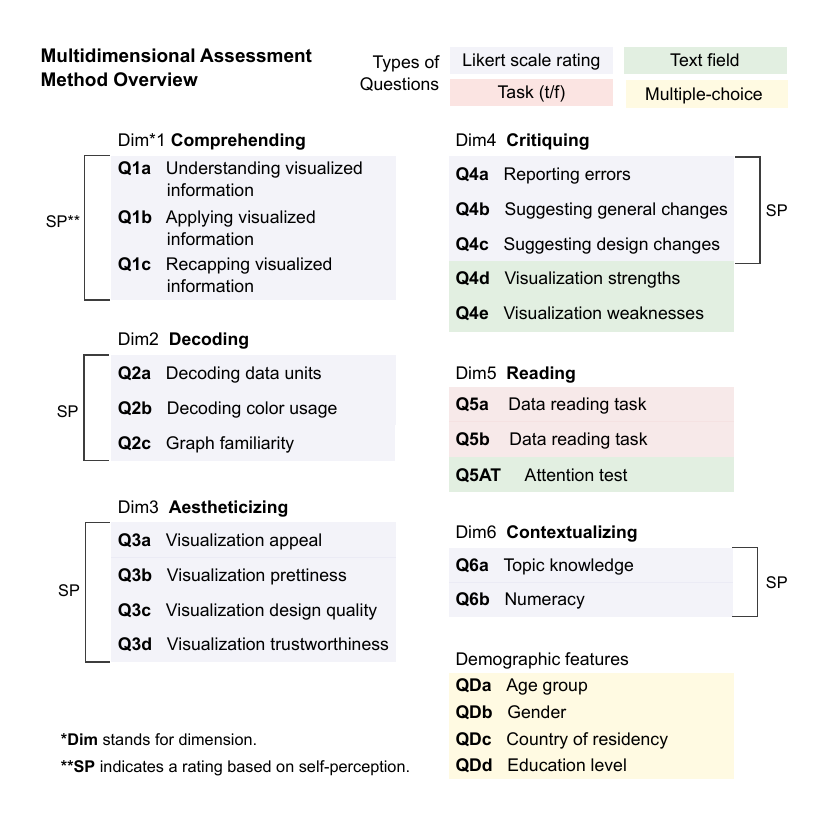}
    \caption{Overview of MdamV's structure with six assessment dimensions (Dim1–Dim6), alongside demographic questions. For the full question catalog, see the supplementary material.}

    \label{fig:overview-batteries-questions}
\end{figure}

With MdamV (Multidimensional Assessment Method), we suggest six assessment dimensions (see Fig. \ref{fig:overview-batteries-questions}) that we constructed for visualization understanding and evaluated in a survey. Four out of six dimensions (\textit{Comprehending}, \textit{Decoding}, \textit{Aestheticizing}, \textit{Contextualizing}) rely on self-perceived ability ratings (see Sec.~\ref{sec:saps}). One dimension (\textit{Reading}) covers low-level visualization tasks, while another offers optional text inputs (\textit{Critiquing}) for open-ended critique. The dimensions are outlined below in the order they were assessed, without implying a hierarchy. Similar to how Wiggins and McTighe~\cite{wiggins2005} describe for their framework on \textit{understanding} (see Sec.~\ref{sec:deconstruct}), we do not claim our proposed dimensions to be the only way of capturing understanding. Instead, we aim to break visualization understanding into different facets, making underexplored abilities in prior assessments (see Sec.~\ref{sec:background-limitations}) more visible.\\ 

\noindent\textbf{Dimension 1: Comprehending} focuses on overall perceived understanding, by addressing general impressions of the provided data visualization (Q1a/b). Further, assuming that abstract thinking is an indicator of a high level of literacy~\cite{bowen2017}, respondents are asked if they could easily summarize the content of the visualization to someone (Q1c).

\noindent\textbf{Dim2: Decoding} considers self-perceived decoding of graph specifics, such as graph elements and familiarity with the displayed chart type. This includes querying perceived comprehensibility of color usage (Q2b) and depicted measurement units (Q2a), which reflects that visualization comprehension can depend on both structural/content-related features and visual/aesthetic design elements~\cite{robinson1984}. Moreover, graph familiarity is said to be beneficial for reading data correctly~\cite{fischer2018}, and therefore, is also investigated (Q2c).

\noindent\textbf{Dim3: Aestheticizing}, we adapted the \emph{BeauVis scale}~\cite{he2023}, which consists of synonymous English-language adjectives aimed at capturing aesthetic perception. The terms were translated into German, the primary language for testing (see Sec.~\ref{sec:quantitativerun}), yielding three items (Q3a-c). To complement these, we added a question on perceived trustworthiness (Q3d), as attributing beauty to a visualization may favorably bias its interpretation~\cite{chujun2021}.

\begin{figure*}[ht]
    \centering
    \begin{subfigure}{0.45\textwidth}
        \centering
        \includegraphics[width=\textwidth]{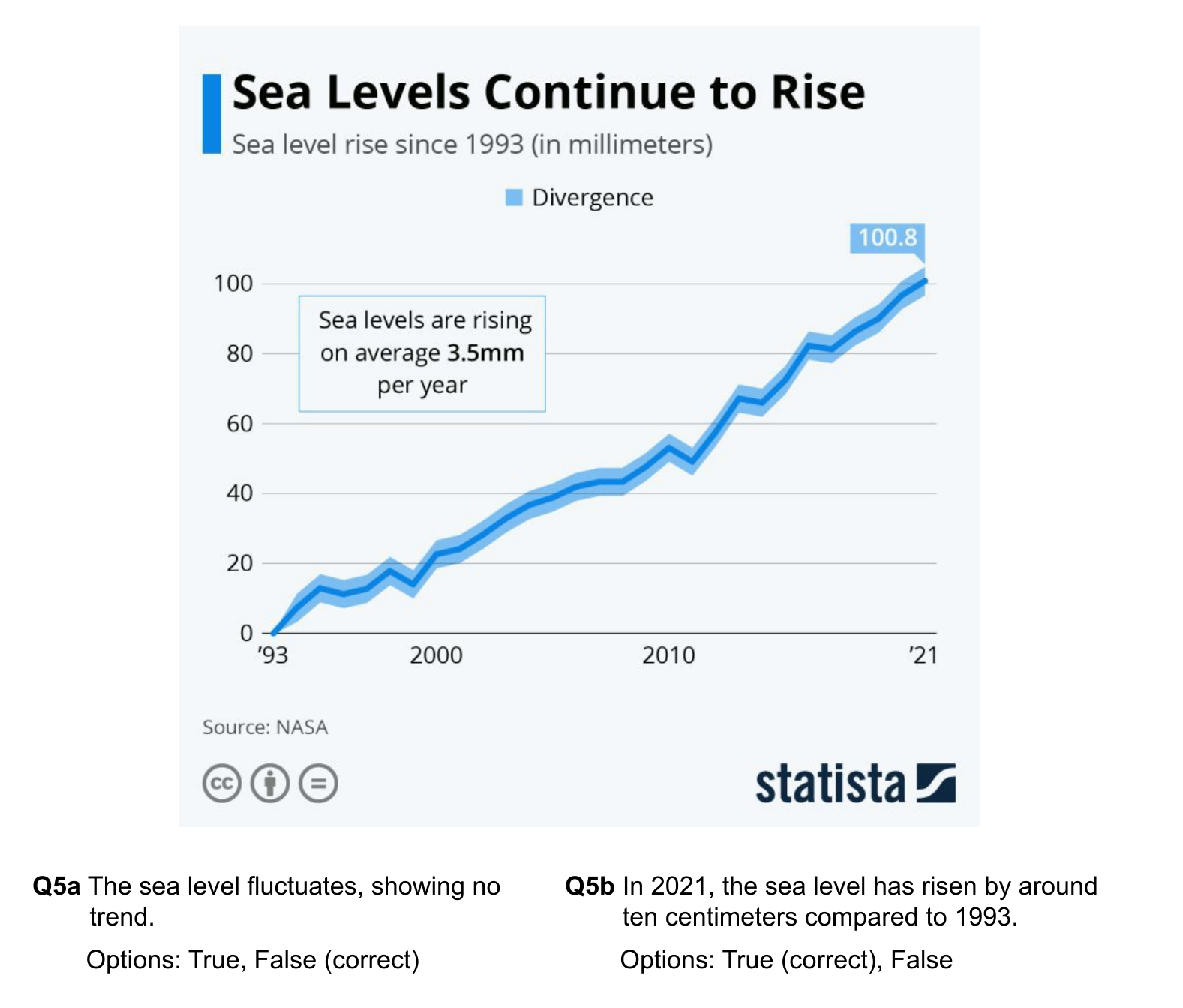}
        \caption{Data Visualization 1: A \textbf{line chart} showing sea level change over time with associated data reading tasks (Q5a, Q5b) and marked correct answers. \label{fig:line-chart}}
    \end{subfigure}
    \hfill
    \begin{subfigure}{0.45\textwidth}
        \centering
        \includegraphics[width=\textwidth]{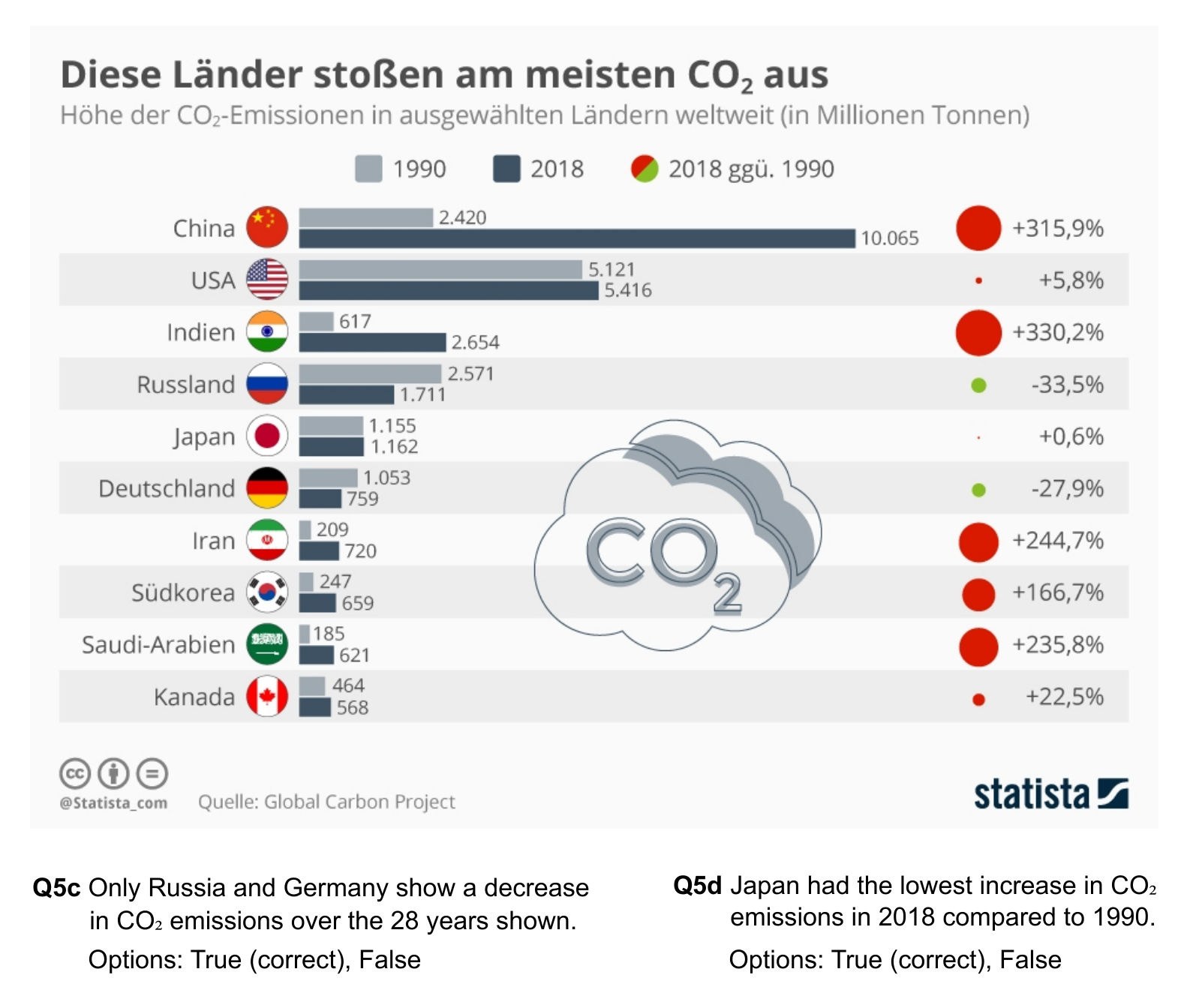}
        \caption{Data Visualization 2: A \textbf{bar chart} comparing $\text{CO}_2$ emissions across selected countries (1990–2018) with tasks (Q5c, Q5d) and indicated correct answers. \label{fig:bar-chart}}
    \end{subfigure}
    \caption{Data visualizations used in the MdamV survey with a German-speaking audience. Figure (a) shows an official English version of the tested line chart, while Figure (b) displays the original bar chart used in the survey, which is only available in German. See Sec.~\ref{climate-data-vis} for details.}
    \label{fig:vis1+vis2}
\end{figure*}

\noindent\textbf{Dim4: Critiquing} centers on critical ability and understanding of visualization construction. Respondents are asked to articulate specific feedback on the displayed data visualization. This includes indicating whether they have any erroneous findings to report (Q4a) or specific amendment proposals to share (Q4b/c), as literacy is also said to cover the ability to question an object or seek out possible changes~\cite{chevalier2018} as well as to create visualizations~\cite{brill2007, ge2025avec}. Respondents were offered subsequent optional text fields asking for visualization strengths and weaknesses (Q4d/e).

\noindent\textbf{Dim5: Reading} uses task-based measures in the form of data reading questions (Q5a/b), a well-established way to assess literacy performance (see Sec.~\ref{sec:background-limitations}). We pose true-or-false questions (see \autoref{fig:vis1+vis2}), which directly reference the data content of the tested visualizations.

\noindent\textbf{Dim6: Contextualizing} covers two context-related facets of visual data literacy: perceived numeracy and topic knowledge. A higher level of numeracy has been shown to increase the likelihood of reading a data visualization correctly \cite{reyna2009, xexakis2021}, which is why we asked respondents to rate their own numeracy level (Q6b). In addition, we asked about personal interest in the visualization's topic (Q6a), as it may influence the understanding of visually represented data~\cite{fischer2018}.

Though most questions are generally phrased, some reference the data visualization example to exemplify parts of the question to respondents. This is the case for data reading tasks (Q5a/b) bound to the visualization content. Specifications were also made to Q2a by listing measurement units and Q2c by naming the chart type. In Q6a, respondents were asked to rate personal interest in the displayed topic.

\section{Methodology}
The design and testing of MdamV were split into four phases. At first, questions were developed in a theory-driven approach (see Sec.~\ref{sec:Question Batteries}) and placed within a digital survey format. Then, in two phases of qualitative testing, the assessment was presented to eight participants. Semi-structured interviews allowed us to gain insights into the validity of MdamV questions, which were followed by iterative development. Finally, the assessment was completed by 438 respondents representing the Austrian population's age groups, from 18 to 74 years, and their male-female gender distribution. MdamV was developed and tested in the country's primary language. The English version presented here is a translation and has not been tested. The original questions are made available at an anonymized link (to be provided in the camera-ready version).

\subsection{MdamV in Practice: Designing a Survey}
MdamV was implemented using \textit{SoSci Survey}, a survey design software free for academic purposes~\cite{sosci}. The assessment was tested with four data visualizations throughout the pilots, and two of them were used in the survey conduction, with each respondent always encountering two visualizations to minimize bias from any single visualization. To prevent order effects~\cite{boy2014}, automated urn-drawings randomized the display sequence. This section discusses the rationale behind our visualization selection for testing and outlines survey tasks, which are fully detailed in the supplementary question catalog.

\subsubsection{Selection of Visualization Examples on Climate Data}
\label{climate-data-vis}

As highlighted in Sec.~\ref{sec:perspective}, considering visualization-centered factors is valid when treating visual data literacy as a situated process, because each visualization can expose specific insights into how readers construct understanding. To ground our study in a relevant context, we focused on climate data visualizations, since they are an essential medium for communicating critical information on the climate discourse~\cite{moser2010communicating, o2014climate, schneider2012climate, kim2017}. While they can be effective tools for conveying complex information~\cite{okan2011, fischer2018}, even the Intergovernmental Panel on Climate Change has acknowledged challenges in making them comprehensible to broader audiences~\cite{ipcc2016,harold2019}. Studying their effectiveness is essential to understanding audience engagement with climate data better and ensuring trust in climate science~\cite{schuster2022}.

The selection of data visualizations for MdamV testing followed a number of criteria: they had to present climate-related data to match our topic setting, be simple enough to encourage engagement, they had to differ in chart type and remain static rather than interactive. Additionally, all visualizations were licensed under Creative Commons (CC BY-ND) for public use and credited accordingly. Their suitability for a general audience was judged during short feedback exchanges with visualization researchers.

Based on these criteria, we selected a line chart and a bar chart (see Fig. \ref{fig:vis1+vis2}), two of the most widely used visualization types~\cite{munzner_visualization_2014, borkin2013massvis}, to ensure that respondents could engage with them regardless of prior experience. 
The \textbf{line chart} (Fig. \ref{fig:line-chart}) shows sea level rise from 1993 to 2021, with an annotation highlighting an average annual increase of $3.5$ millimeters. The line is shaded with a light blue area labeled "Divergence". The \textbf{bar chart} (Fig. \ref{fig:bar-chart}) compares $\text{CO}_2$ emissions of $10$ countries in 1990 and 2018, using color-coded bars and percentage changes. Each country's flag is displayed, and a $\text{CO}_2$ cloud icon is placed in the middle of the chart. During testing, both charts were in German, as this was the primary language of the surveyed audience.

\subsubsection{Survey Tasks}
During testing, respondents were shown two data visualizations, so some assessment dimensions (Dim1-4) appeared twice, once for each visualization. Questions on \textit{Contextualizing} (Dim6) appeared only once. \textit{Reading} questions (Dim5) featured four tasks in true-or-false mode, two per visualization. 
An additional reading task for the bar chart, which had to be answered in an obligatory text field, served as an attention check to ensure survey data quality.

To cater to a general audience, there were no prerequisites for taking the survey. Respondents were first informed about the study's purpose of investigating the understanding of climate data visualizations. They had to confirm their consent to participate in the study and that they were not underage. 
Participants were asked to engage with two data visualizations, focusing on both content and overall impression. After viewing each visualization, they proceeded to a set of related questions. Whenever a question referred directly to a visualization, the corresponding chart was displayed beneath it. Self-perceived ability ratings were presented on a five-point scale ($1$ = I disagree, $2$ = I somewhat disagree, $3$ = Undecided, $4$ = I somewhat agree, $5$ = I agree). For analysis, ratings of $1$ and $2$ were grouped to represent disagreement, and $4$ and $5$ as agreement. A rating of $3$ was considered a neutral answer. In Dimension 4 on \textit{Critiquing}, respondents could provide optional text input, which was evaluated qualitatively by grouping responses into categories.


\subsection{Qualitative Testing: Survey Pilots}
\label{sec:pilots}

The survey was piloted qualitatively with eight participants, who completed MdamV questions on two visualizations in the presence of a researcher. In semi-structured interviews, participants discussed their survey experience and explained how they interpreted the tested data visualizations.

\subsubsection{Purpose and Procedure of Survey Pilots}
The survey pilots helped refine and validate MdamV questions by qualitatively evaluating respondents' experience and iterating on question clarity and feasibility. The insights were particularly productive for fine-tuning question phrasing. Respondents were recruited purposively through professional and personal relations, aiming for diverse educational and professional backgrounds. Six out of eight respondents had no expertise in climate data visualizations, one respondent was familiar with data visualizations due to their field of work, and one was a visualization researcher. The survey pilots were split into two phases, each consisting of four pilots. Sessions lasted between 28 and 47 minutes (median: 35 min). Interviews were audio-recorded with the consent of all participants and then transcribed.

\subsubsection{Analysis of Survey Pilots and Learning Excerpts}
The interview data was analyzed thematically by accumulating interview passages into themes and deriving a subsequent interpretation for assessment development. The analytical process took place in reference to Grounded Theory~\cite{strauss2015}. This qualitative approach combines interview data into bundled theoretical interpretations justified by the subject matter. Interview data was coded and then assigned to categories~\cite{bryant2010} to gain an overview of respondents' experience. The following two themes are exemplary as categories used for further survey development, which are based on the coded interview data:

\noindent\textbf{Issues of Question Comprehensibility.}
In the first pilot phase, all respondents noted unclear phrasing, with some questions requiring rereading, which led to targeted revisions. In the second phase, only two respondents mentioned unclear phrasing, and rereading only occurred in just one case. 

\noindent\textbf{Engagement With Text Fields.} \textit{Critiquing} included two optional text fields per data visualization. Participants' willingness to use them varied, reflecting different attitudes towards expression in text form. Three out of eight respondents principally skipped text fields; two articulated some thought; another two did not fill in the fields, even though they verbally expressed suitable inputs to the present researcher. The eighth respondent filled out all text fields conscientiously, as was self-stated, to perform well in the survey. Since pilot evaluation showed that respondents varied in how they expressed themselves, to avoid bias from differences in expressive ability or response attitude, text inputs for assessing critical ability were kept optional. 

\subsection{Quantitative Testing: Representative Survey}
\label{sec:quantitativerun}
After piloting, we refined MdamV's questions and survey implementation to collect data from a panel representative of the Austrian population's age groups (18-74 years) and male-female gender distribution within those age groups (N=$438$).

\subsubsection{Data Collection With a Market Research Organization}
\label{coop}
So that MdamV could undergo an application with a representative scope, respondents were recruited in cooperation with a market research organization. The organization guided respondents to the survey in a controlled manner and pre-selected them based on crossing the agreed-upon quotas of age groups and gender. The respondents originated from the organization's database or other websites promoting market research participation. Respondents received incentives, whose form depended on the panelist's affiliation with either the market research organization or other platforms. The organization held demographic self-reported data from the panel, the accuracy of which was verified by collecting demographic data in our survey.

During a period of two weeks, respondents were directed to our online survey, with survey cases achieved in subsets of 50 completions. We evaluated the data quality of survey completions after each subset, followed by the criteria for valid survey cases, as illustrated in the next subsection.

\subsubsection{Quality Criteria for Valid Survey Cases}
\label{sec:criteria}
The completed survey cases were downloaded as a data set from \textit{SoSci}, where the survey had been self-administered on our part. All 457 cases from the market research cooperation period from March 3 until March 17, 2023, were selected. A completed case held responses for all obligatory questions (i.e., excluding the optional text fields), and both tested data visualizations. Respondents who exited the survey early were omitted. Another 19 cases were disregarded, as these respondents failed to answer the attention question (see Q5AT in Fig. \ref{fig:overview-batteries-questions}) correctly. Referring to the surveyed bar chart (Fig. \ref{fig:bar-chart}), this question asked respondents to name the first country listed in the $\text{CO}_2$ emissions chart. The correct answer was China, with variations like "china" or "Chin" also accepted. The resulting data set, with 438 cases that passed the attention test, will be made openly accessible upon publication of this paper.

\subsubsection{Demographics of the Survey}
\label{demographics}
Demographic information on the respondents included their gender, age group, and education level. Respondents had to confirm their country of residence where the survey was conducted. Inclusive options for gender self-assignment were offered. Of the $438$ respondents, $223$ were male ($50.9\%$) and 214 ($48.9\%$) female. One respondent reported their gender as inter. Age groups were recorded in categories of 10-year ranges and represented the distribution of the surveyed population. The minimum registered age was 18 since the legal age was a consent criterion, and the maximum was 74 years, as this was the highest age of online panelists available to the market research organization.

The representation of education levels, though meaningful, was not sought after because, among other things, the standardized procedure of the market research organization did not provide for the representation of this demographic feature. The sample turned out to over-represent respondents with higher education, such as university degrees.

\section{Results}
\label{sec:results}
We tested MdamV with $438$ respondents, representative of Austria's population in terms of age groups (18-74 years) and male-female gender split. Using MdamV on a line and a bar chart of climate data (see Fig. \ref{fig:vis1+vis2}), we gathered insights into respondents' visualization understanding. The following results illustrate the feasibility of our proposed method for assessing situated visualization understanding.

\begin{figure}[h]
    \centering
    \includegraphics[width=0.9\columnwidth]{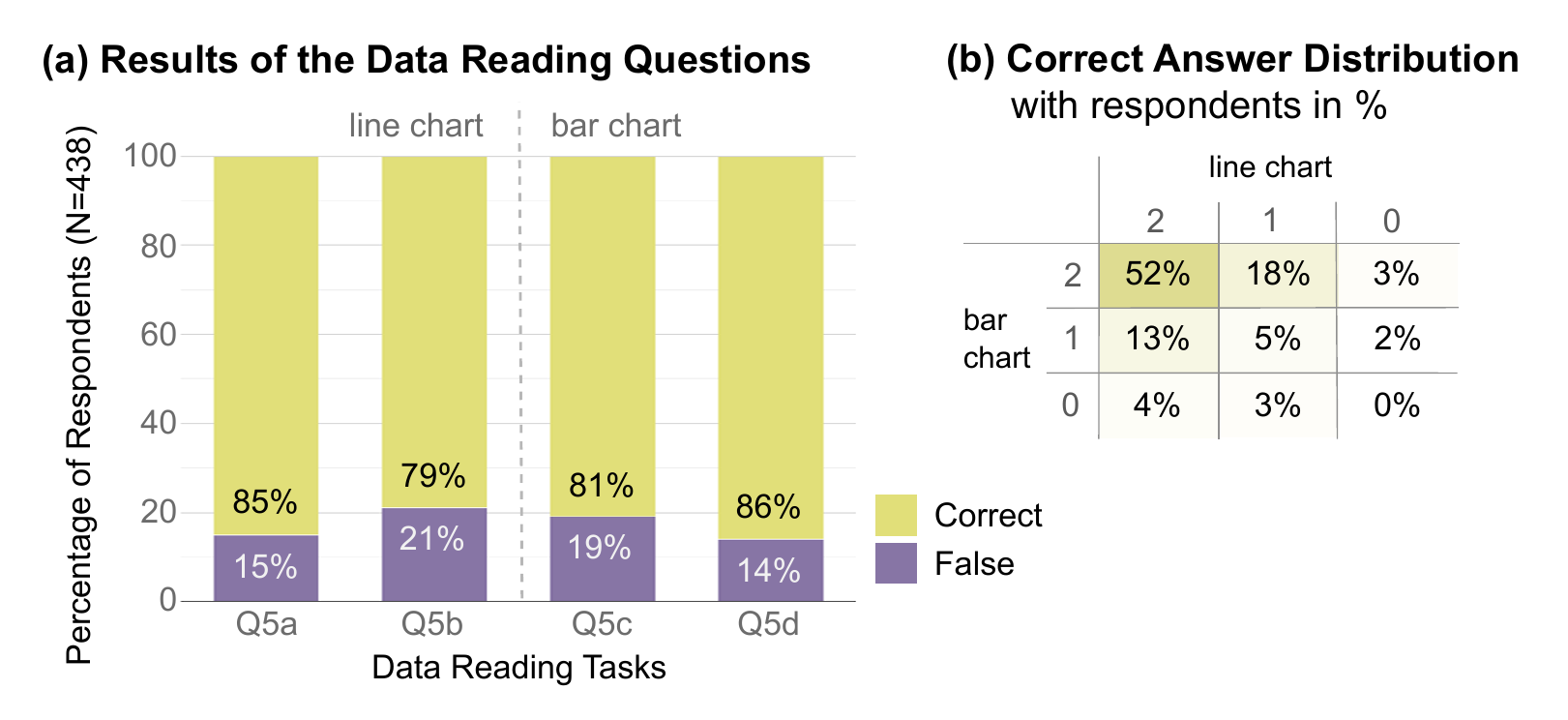}
    \caption{Data reading tasks (Q5a-Q5d) were presented in true-or-false mode (see Fig. \ref{fig:vis1+vis2}). (a) shows correct vs. false responses and (b) the distribution of correct answers per chart, both with N=438.}
    \label{fig:data-reading-results}
\end{figure}

\subsection{Insights into Performance Measures}

To give a general impression about completing the assessment: It took respondents an average of $2.75$ minutes to answer the main body of MdamV questions per data visualization. On average, framing elements of the survey, such as the introduction and the query of demographic info, took about a minute to complete.

Based on the \textit{Reading} questions, there were clear deficits when evaluating the surveyed population's visualization understanding from a task-based point of view. Overall, $48\%$ of the respondents made mistakes when reading the simple line and bar chart. Effect sizes of differences between the two charts are relatively small, as seen when comparing the answer distribution of correct answers (see Fig. \ref{fig:data-reading-results}a). Just over half of respondents solved all four data reading tasks correctly, while none answered all of them incorrectly (see Fig. \ref{fig:data-reading-results}b).

\subsection{Respondents' Perception of the Surveyed Charts}
\label{sec:results-selfassessed}

In contrast to the data reading performance, a considerably smaller portion of the respondents indicated that they are not able to summarize the content of the shown data visualizations (see Q1c in Fig. \ref{fig:overview-batteries-questions}). In the case of both charts, only $5\%$ of respondents stated that they cannot summarize the visualization. This suggests the possibility of misleading information transfer, as many respondents are unaware of reading deficits. Regarding the displayed chart types (Q2c),
The intersection of these statements results in $12\%$ of respondents who felt unfamiliar or uncertain with the line and the bar chart as chart types. Further, respondents have self-assessed deficits in understanding simple measurement units (Q2a). $22\%$ of respondents did not fully comprehend the usage of millimeters, and $28\%$ of respondents found millions of tons somewhat incomprehensible.

There was a notable difference in the aesthetic perception of the surveyed climate data visualizations. Around $10\%$ of respondents rated the impression of the bar chart more positively than that of the line chart. In particular, $60\%$ agreed to find the surveyed bar chart pretty (Q3b), and $72\%$ found it well-designed (Q3c). The line chart was perceived as slightly less aesthetic, with $58\%$ agreeing to find it pretty and $68\%$ finding it well-designed. This information tends to agree with the assessment of visualization trustworthiness (Q3d), as $75\%$ of respondents found the bar chart trustworthy, with a negative five percent difference from those who found the line chart trustworthy.

\subsection{Correlative Exploration of MdamV Variables}
\label{sec:correlation}
An exploratory look at correlations between MdamV variables supports the distinction of assessment dimensions that we proposed in Sec.~\ref{sec:Question Batteries}. For example, MdamV questions within Dimension 1-4 show steady inner cohesion and simultaneously, these dimensions differ from one another, suggesting they query distinctive ability areas (see Sec.~\ref{subsec:corr}). The correlative exploration highlights some variables, such as self-assessed numeracy, as significant indicators for visualization understanding in terms of correct data reading (see Sec.~\ref{subsec:significance}). 

Statistical analyses were conducted using Python (scipy.stats) and R (MASS, psych). Alongside non-parametric tests, we also used parametric methods, following arguments that they are appropriate for Likert-scale data even with small sample sizes, unequal variances, or non-normal distributions~\cite{NormanGeoff2010}. The correlations (in Sec.~\ref{subsec:corr}) were calculated according to Pearson’s correlation coefficient using the median whenever results had to be aggregated over dimensions or visualizations. To cross-check and thus confirm our statistical results, we calculated Spearman's rank correlation coefficient for the questions within each dimension and found that the results were almost identical, with deviations between the two correlation coefficients below $0.05$. For binary results (in Sec.~\ref{subsec:significance}), a t-test for the arithmetic mean and Mood's median test were conducted and gave similar results. Both tests gave the same result with regard to rejecting the null hypotheses.
Additionally, we looked at ordinal regression, with the outcomes supporting the hypotheses.

\begin{figure}[ht]
    \centering
    \includegraphics[width=0.8\columnwidth]{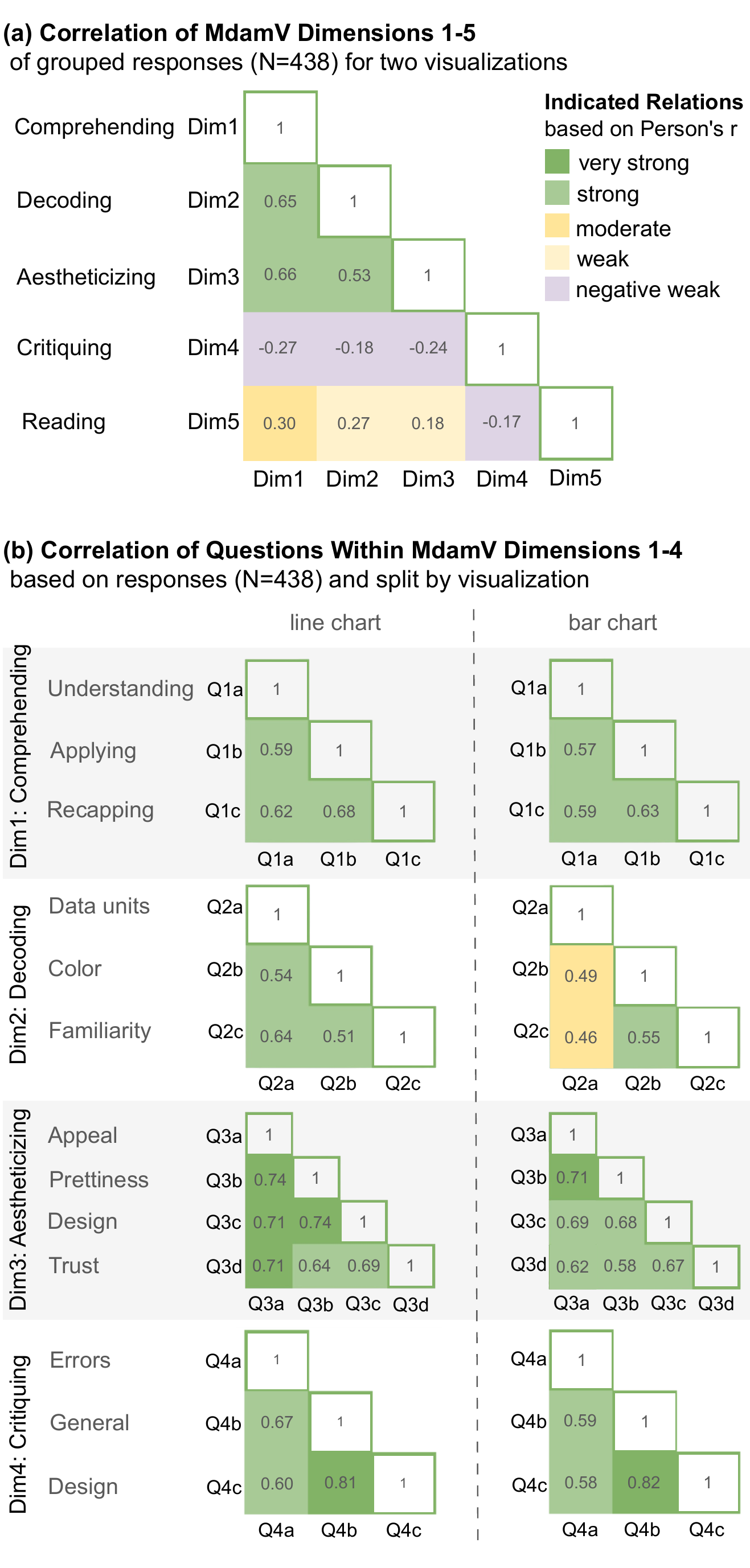}
    \caption{Correlation tables for (a) MdamV Dimension 1 to 5 and (b) questions within Dimensions 1 to 4 per surveyed chart. Indicated relations are based on Pearson's r, and all correlations are statistically significant ($p < 0.05$). See supplementary Fig.~S1–S3 for more detailed views.}
    \label{fig:corr}
\end{figure}

\begin{figure*}[ht]
    \centering
        \begin{subfigure}{0.45\textwidth}
        \centering
        \includegraphics[width=\textwidth]{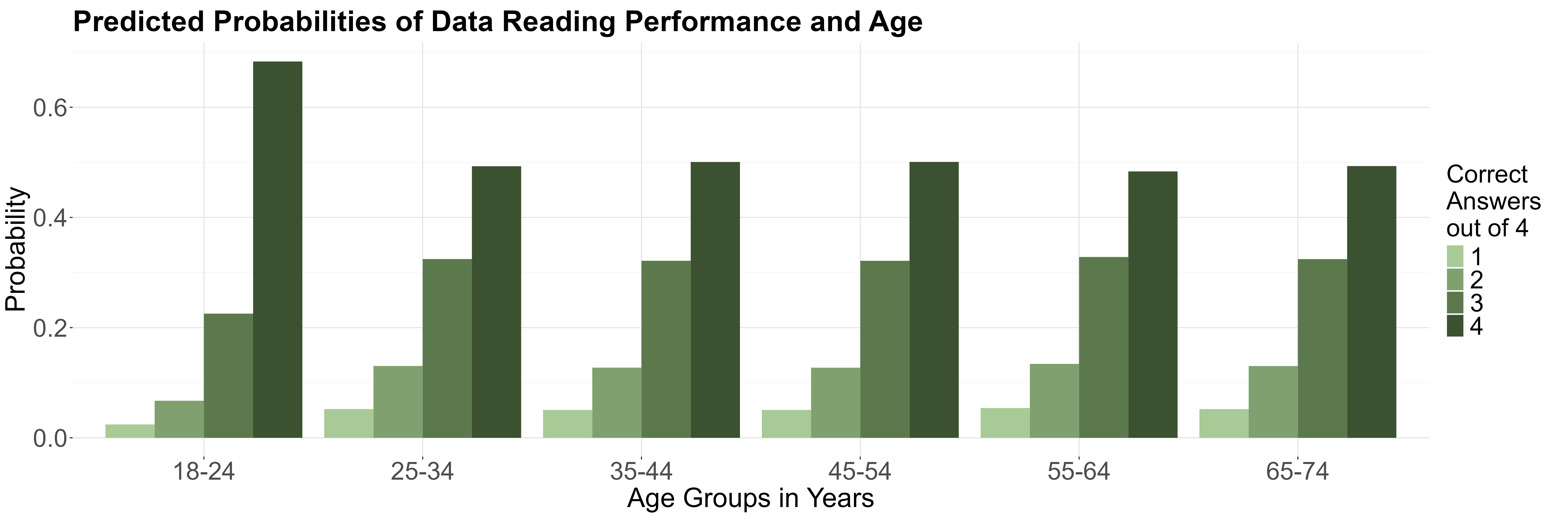}
        \caption{Probability for each data reading score depending on the age category of the respondents. Younger respondents had a higher probability of answering all data reading tasks correctly.}
        \label{fig:ordinal-regression-age}
        \end{subfigure}
    \hfill
    \begin{subfigure}{0.45\textwidth}
        \centering
        \includegraphics[width=\textwidth]{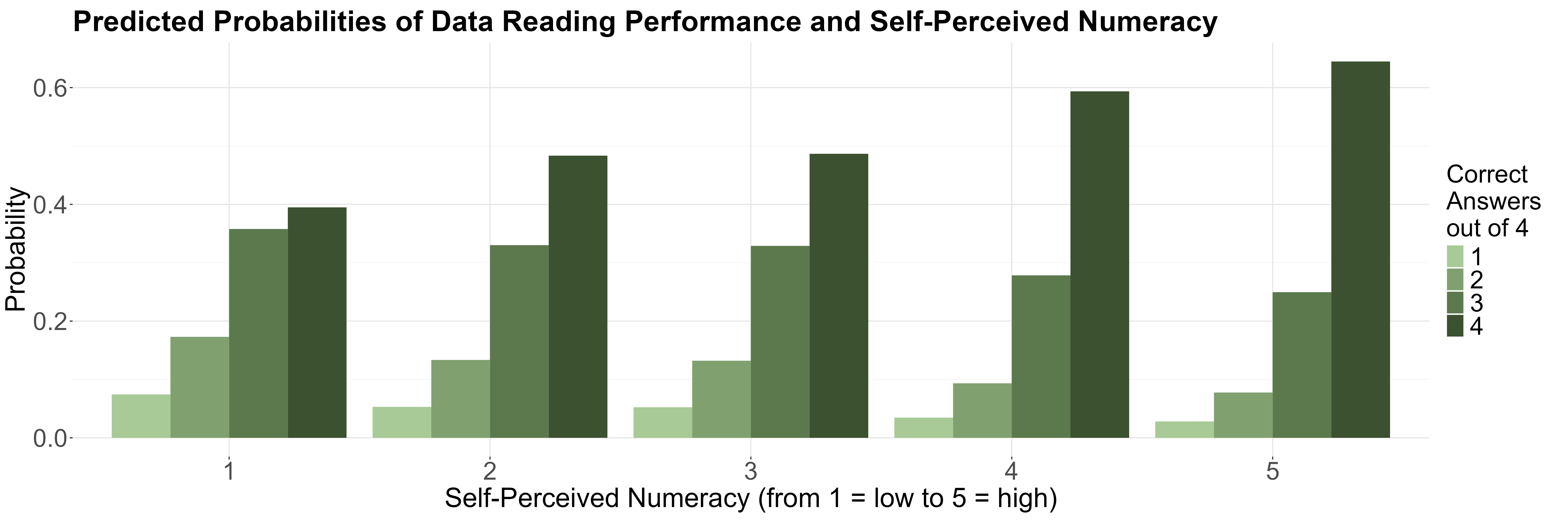}
         \caption{Probability for each data reading score depending on the numeracy self-assessment of the respondents. Higher self-rated numeracy increased the likelihood of correctly answering all data reading tasks.}
    \label{fig:ordinal-regression-num}
    \end{subfigure}
    \caption{Ordinal regression supports the observed relationship between age, self-assessed numeracy, and data reading task performance.}
    \label{fig:ordinal regression}
\end{figure*}

\subsubsection{Correlating Assessment Variables from MdamV}
\label{subsec:corr}

This part of the correlative analysis explores relationships between dimensions as well as between individual questions within dimensions, using statistical methods and significance testing. A consistent positive correlation of questions within Dimensions 1 to 4 indicates the plausibility of constructing assessment dimensions. This holds true for both Pearson and Spearman correlation coefficients. As shown in the correlation tables in \autoref{fig:corr}b based on Pearson's r, the relations between questions within dimensions for both surveyed data visualizations are fairly strong. For questions on \textit{Comprehending} and \textit{Decoding}, all coefficients fall within the moderate (0.3-0.49) to strong (0.5-0.69) range. For questions on \textit{Aestheticizing} and \textit{Critiquing}, correlations are strong to very strong (0.7-1). Note that, for \textit{Critiquing}, only the questions of Likert type (Q4a-c) were considered for the correlation and optional text fields are evaluated separately (see Sec.~\ref{report-viscriticism}). 

To evaluate internal consistency, we also calculated Cronbach’s alpha for Dimensions 1 to 4, which ranged from $0.75$ to $0.9$, indicating good to excellent reliability. These results further support the aggregation of items within each dimension. Given the distributional properties of the data, we opted to use the median as a robust summary measure. 

When dimensions were fully grouped across questions and visualizations, as done in \autoref{fig:corr}a, \textit{Comprehending} strongly correlates with \textit{Decoding} and \textit{Aestheticizing}. Their correlation with the \textit{Reading} performance varies and remains weak (0.1-0.29) to moderate. \textit{Critiquing} disrupts this dynamic, as it is the only dimension correlating negatively with the other ones. This could be due to the simplicity of the data visualizations, as they might limit the scope of finding errors (Q4a) or coming up with design suggestions (Q4c). Respondents might react differently if hidden errors or intentional design issues were to be part of the assessed visualizations.

\subsubsection{Significance of Age and Numeracy}
\label{subsec:significance}
Age and numeracy correlated significantly with \textit{Reading} performance. To test this, binary t-tests and Mood's median test were applied to MdamV variables, including demographic features, in relation to the task-based reading questions. These responses were then aggregated into a score, summarizing the number of correct answers out of four reading questions. Testing showed that the youngest age group of the surveyed population outperformed the older respondents regarding their data reading performance. The age group from 18 to 24 years ($14\%$ of respondents) had a score median of 4 out of 4 correct answers (with p=0.01). The age groups from 25 to 74 years ($86\%$ of respondents) had a lower score median equal to 3 correct answers (with p=0.01). In addition, the self-assessed numeracy level (Q6b) correlates significantly with the data reading score. Those respondents who rated their numeracy level high ($37\%$ of respondents) tended to have a median score equal to 4 (with p=0.009). Those who rated their numeracy level neutral or low ($63\%$ of respondents) had a lower median score equal to 3 (with p=0.009). This suggests a clear value of having respondents self-assess their ability level.

We further validated these results using ordinal regression. The outcomes showed a substantially higher probability of achieving a data reading score of 4 out of 4 in the youngest age group compared to other age groups (see Fig. \ref{fig:ordinal-regression-age}). Additionally, the regression confirmed the influence of self-assessed numeracy on data reading performance, as shown in the predicted probabilities in Figure~\ref{fig:ordinal-regression-num}.

There is no correlation between the highest level of education reported and the self-assessed level of numeracy (Pearson's r=0.016). Further, as expected, our analysis did not observe a meaningful difference in the provided survey answers based on gender.

\subsection{Report on Visualization Critique in the Text Inputs}
\label{report-viscriticism}
For each visualization, respondents had the option to provide a written visualization critique in two text fields (max. $2,000$ characters each), one for visualization strengths (Q4d) and one for visualization weaknesses (Q4e). These questions were not mandatory, yet $37\%$ submitted at least one meaningful response. All inputs for the line chart and the bar chart were categorized into i. Visualization Strengths, ii. Visualization Weaknesses and iii. Climate Commentary.

\subsubsection{Categories of Text Inputs and Examples}

\textbf{i. Visualization Strengths} were defined as a category of text inputs where a clear positive point of critique related to the tested visualization was made. For the line chart (see Fig. \ref{fig:line-chart}) $99$ inputs were given, such as finding the chart "inviting, easy to understand, the color blue well chosen", or noting it was "kept quite simple not cluttered". Someone else highlighted that "the source is indicated, [which] must be highly credited". For the bar chart (see Fig. \ref{fig:bar-chart}) $113$ respective inputs were provided, describing it as "easy to understand and memorize", acknowledging that it "raises awareness", or finding its "data easily comparable".

\textbf{ii. Visualization Weaknesses} were defined as comments highlighting clear negative points of critique related to the tested visualization. For the line chart, $87$ inputs were provided, stating, for example, that "the graphic looks outdated, could be designed better". A few respondents were "missing axis labeling" or wondered what the uncertainty margin means. For the bar chart, there were $78$ inputs, including debate on the cloud icon, finding it "unnecessary" or "strange" (though this graph element had also been highlighted as a strength, e.g., someone said that the "cloud speaks to [them]"). Others criticized the comparison of countries, and, for instance, found the "relation to population missing" or wanted more information because "there are other countries worth mentioning that also emit high levels of CO2".

\textbf{iii. Climate Commentary} appeared in Q4d and Q4e and emerged in our analysis as opinions about and reactions to climate change, rather than comments on the tested visualization itself. In total, there were $35$ of such inputs. To give an impression of this category, one respondent wrote that "the so-called man-made climate change is pure fraud", others called climate change "a lie" and one respondent said that the visualization would aim towards "tricking people into fear". Others articulated feeling affected, such as that the visualization "makes [them] sad" or that looking at it, they felt that "the world will be destroyed".

\subsubsection{Statistical Analysis of Text Inputs}
\label{statistics-text}

A chi-squared test was employed to evaluate the relationship between the non-ordinal text inputs and demographic factors and respondents' data reading performance, which was summarized as a score. For this test, i. Visualization Strengths ($212$ inputs total) and ii. Visualization Weaknesses ($165$ inputs total) were considered equally by coding them as $1$s and putting them in a 2x2 contingency table for each data visualization. Due to partially small sample sizes, significant findings from the chi-squared test were validated through an additional Monte Carlo simulation~\cite{eth-chisq, shije2010} by randomly sampling $2,000$ replicates. 

For both the line and bar chart, results indicated a significant difference in text input provision across education levels (line chart: ${X}^2$=12.93, df=1, p=0.0003, bar chart: ${X}^2$=7.88, df=1, p=0.005). The simulated p-values, based on $2,000$ replicates, validate the significance of this finding (line chart: p=0.0005, bar chart: p=0.005). This suggests that respondents with secondary and tertiary educational degrees were the ones who tended to provide optional visualization critique. As educational disparities can impact survey responses~\cite{demarest2013}, higher-educated respondents may be more accustomed to survey items demanding written input and thus have a structural propensity to provide this type of answer.

Moreover, a significant association was found between the provided text inputs and the data reading performance (line chart: ${X}^2$=10.40, df=1, p=0.001, bar chart: ${X}^2$=8.91, df=1, p=0.003), and validated through a simulation with $2,000$ replicates (line chart: p=0.001, bar chart=0.002). These findings indicate that the respondents with a higher data reading score of 3 or 4 correct answers (out of 4) were the ones who were more likely to articulate a visualization critique. This underscores that critical ability can be linked to literacy levels~\cite{chevalier2018}. In the chi-squared tests, no significant differences were found regarding the association of text input provision across neither age groups nor gender.

For the subset of 20 respondents who provided inputs categorized as Climate Commentary (35 inputs in total), we applied a binary coding scheme (Climate Commentary: yes/ no) and conducted an independent samples t-test on the overall dataset. The results showed a mean data reading score of $3.32$ for those without climate commentary and $2.9$ for those who shared an opinion; however, they were not statistically significant ($p = 0.097$). While the full-sample t-test results are almost statistically significant, this appears to be driven by the imbalance in group sizes. When accounting for this imbalance through a bootstrap-like approach, the effect was not robust ($p \approx 0.23$), suggesting that the observed difference may be sensitive to sample composition rather than reflecting a consistent underlying difference. Given the small and uneven sample sizes, respondents' attitudes toward climate change may warrant closer examination in future research. 

\section{Discussion}
With MdamV, we propose a broader lens for assessing visualization understanding: one that acknowledges performance, while integrating self-perceived ability ratings and voluntary written critique. Our aim is not to replace instruments like VLAT~\cite{kim2017} or CALVI~\cite{ge2023}, but to broaden their performance-focused approach by integrating additional facets of visualization understanding into a more multidimensional method. In our survey, we applied MdamV to two climate data visualizations and gathered insights into how general audiences not only read but also perceive and reflect on their engagement with the tested data visualizations. As displayed in our results (see Sec.~\ref{sec:results}), these are insights that task-based tests alone cannot fully capture. 

In this section, we reflect on the conceptual limitations of our assessment design and comment on survey-related limitations. Further, we discuss what our survey results reveal about the different assessment dimensions for visualization understanding that we proposed, and outline directions for expanding MdamV's applicability. 

\subsection{Survey-Based Insights from MdamV}


Drawing on frameworks that deconstruct understanding into ability facets (see Sec.~\ref{sec:deconstruct}) and prior work related to visual data literacy (see Sec.~\ref{sec:perspective}, Sec.~\ref{sec:Question Batteries}), we constructed six ability dimensions for assessing visualization understanding, bundled in a method we call MdamV. 
We collected insights not only on performance through data reading (MdamV Dimension 5) but also on perceived comprehension (Dim1) and decoding ability (Dim2), aesthetic judgment (Dim3), critique behavior (Dim4), and context-related self-assessments for numeracy and topic interest (Dim6). While our findings are tied to the visualizations tested, they can offer valuable insights for visualization design practices and the communication of critical data to general audiences.

Our survey results show that even with some of the most common chart types~\cite{borkin2013massvis, munzner_visualization_2014}, a notable share of respondents indicated unfamiliarity with them, which is exemplified by an intersection of $12\%$ of respondents feeling unfamiliar with \textit{both} line and bar charts (see Sec.~\ref{sec:results-selfassessed}). About a quarter of respondents felt unable to decode data units like millimeters and millions of tons, highlighting that elements implicitly assumed to be self-explanatory by designers or curators~\cite{kolko2010implicit} may require reconsideration.
This points out that correct interpretation by general audiences cannot be taken for granted, underscoring the value of identifying friction points in audience understanding, which can in turn inform visualization design and implementation~\cite{koesten2025crisis}. Such findings also call into question one-size-fits-all approaches, which have been criticized for being ineffective with diverse users who have different literacy levels and types of knowledge~\cite{kostelnick_cartographic_2013,skinner_physicians_1994}.

Including \textit{Aestheticizing} in our assessment revealed that respondents who found a visualization visually appealing also tended to trust it (see Sec.~\ref{sec:results-selfassessed}), suggesting that, in line with prior work~\cite{dork2013criticalvis}, design and perceived credibility are closely linked. This underscores the value of investigating which visual elements or design choices different audience groups may perceive as appealing or trustworthy. For example, proximity techniques have been shown to increase viewer engagement~\cite{campbell2019feeling, kennedy2018}, and such insights could guide the creation of visualizations that are both inviting and credible. Our survey responses reflected this, since inputs on visualization critique pointed to reaction toward specific embellishment, such as to a $\text{CO}_2$ cloud icon (see Sec.~\ref{report-viscriticism}). Although respondents expressed clear preferences, e.g., overall favoring the tested bar chart over the line chart (see Sec.~\ref{sec:results-selfassessed}), aesthetic perception showed only weak correlations with reading performance (see Fig.~\ref{fig:corr}a), indicating that aesthetics had a limited impact on understanding.

Respondents' open-ended critiques (see Sec.~\ref{report-viscriticism}) demonstrated critical engagement and provided valuable qualitative insight into their understanding of the tested visualizations. While such inputs cannot fully capture the complexities of the reader's critical ability, they nonetheless indicate perspective-taking. In Wiggins and McTighe's framework~\cite{wiggins2005}, perspective is said to require a demonstrable shift in understanding. Although we did not test for this directly, we found that when readers step outside their role and reflect on strengths or suggest alternative design choices, they effectively adopt a designer’s viewpoint, which could represent such a shift. 

In \textit{Critiquing}, respondents provided a total of $377$ points of critique on visualizations' strengths and weaknesses, with some comments highlighting distracting design choices or missing labels, offering grounded insight into what viewers notice and what may hinder understanding (see Sec.~\ref{report-viscriticism}). Yet, in their self-ability ratings, most respondents did not feel confident suggesting design improvements (see Sec.~\ref{subsec:corr}), which may indicate a gap between recognizing critique and articulating it. Possibly, more flawed or intentionally misleading visuals (for instance, done by Camba et al.~\cite{camba2022deception} or Ge et al.~\cite{ge2023}) would elicit stronger rating responses, and future applications could explore this space.

\subsection{Limitations}

The integration of self-perceived ability ratings in MdamV offers valuable insights into how people engage with visualizations beyond task-based performance, but it may potentially introduce sources of noise. Self-assessments can be influenced by factors like over- or underestimation, social desirability, or ambiguity in question interpretation~\cite{greven2009spas, mattheos2004selfassessment}. While we have conducted pilots in several qualitative rounds to ensure clarity and ease of interpretation (see Sec.~\ref{sec:pilots}), biases inherent to self-assessment may remain. Nevertheless, as discussed in the Background (see Sec.~\ref{sec:saps}), self-perceived ability ratings are a recognized approach in the study of abilities and continue to provide valuable insights~\cite{hamilton2007, greven2009spas, kritzinger2021trust}.

Regarding our survey conduction, there are some representation boundaries. Respondents below the age of $18$ and above the age of $74$ were not considered in the survey because of legal age as a consent criterion and due to availability reasons (see Sec.~\ref{demographics}). The respondents share the commonality of being incentivized by a market research organization or related platforms, which may introduce a bias toward individuals with frequent online survey participation and, therefore, higher digital skills. However, the approach we have chosen to select respondents is also standard in other representative surveys (e.g.,~\cite{cannizzaro2020, hajek2022, cates2015}). We recognize that respondents' processes of understanding are likely to vary based on their socio-cultural and geographical backgrounds, limiting the scalability and comparability of our results. As emphasized throughout, MdamV was tested with a specific panel and can only be interpreted in relation to the visualizations used. 

Since we investigated \textit{situated} visualization understanding, our survey data, and therefore our results, are tied to the tested visualizations. We selected a static line chart and bar chart for their assumed broad familiarity (see Sec.~\ref{climate-data-vis}) to keep the assessment relevant to general audiences. Still, two chart types cannot represent the variety, complexity levels, and design possibilities of data visualizations. 
Despite this focus, MdamV enabled us to capture and discuss meaningful facets of visualization understanding that may remain hidden even in assessments with more diverse chart sets, such as the Mini-VLAT~\cite{pandey2023}, which relies on a single item per chart type.

We deliberately selected climate visualizations for their real-world relevance, aiming to assess how people engage with meaningful, high-stakes information rather than artificial stimuli (see Sec.~\ref{climate-data-vis}). While such topics may evoke emotionally or politically charged reactions, we argue that these responses are a valid part of the understanding process and reflect how visualizations are encountered in practice. Studying potentially polarizing visuals is, therefore, not only justified but essential for understanding real-world information retrieval and addressing misinformation~\cite{lisnic_misleading_2022, ferrara2020misinformation}. To account for possible bias, we let respondents assess their topic interest (see Q6a in Fig.~\ref{fig:overview-batteries-questions}), just as we examined unsolicited climate-related commentary in our qualitative data and tested whether these participants differed in their data reading performance, finding no significant effect (see Sec.~\ref{statistics-text}). That said, we acknowledge that topic-related influences, whether related to climate or other subjects, would benefit from further development. This could mean including additional questions on topic interest to provide more nuanced contextualization of responses in future surveys.

\subsection{Future Work}
\label{sec:future}

Future applications of MdamV should be expanded to include a wider variety of visualization formats. For example, beyond static charts, interactive visualizations are increasingly common in online communication. They are particularly relevant for further developing our assessment method, since they may require abilities distinct to digital interaction that shape audience engagement~\cite{jia2024interactive,firat2022interactive}. Other common visualization types, including infographics, could be explored to reflect broad visual data communication practices~\cite{lazard2015, huang2019, haddow2014, hahn2021}.

MdamV could further benefit from integration with existing approaches like Mini-VLAT~\cite{pandey2023}, as a combined assessment would allow for further comparison of self-assessed and task-based results within the same sample. Initial correlations, such as between self-assessed numeracy and data reading performance (see Sec.~\ref{subsec:significance}), highlighted promising grounds for such work. For the evaluation of our method, we focused on visualized climate data. However, MdamV could be adapted to other domains by rephrasing topic specificity in the \textit{Contextualizing} dimension, offering a promising direction for future work on assessing its transferability.

Another direction is to make the approach more generalizable by adapting it to study setups that address different phenomena in visualization research. This is particularly relevant where user studies have been critiqued for focusing too narrowly on performance measures~\cite{figueiras2018performance}, lacking qualitative insights~\cite{hyk2025qualquant}, and context-related factors~\cite{hall2022performance}. For example, a "MdamT" could be developed to assess the situated understanding of visualization tools by examining how users, e.g., perceive their ability to work with the tool and relate to its features. Similarly, in the context of XAI, "MdamX" could offer an approach to how users, e.g., understand and interpret specific explanations for AI systems. Such adaptations may yield insights that directly inform tool or system design and refinement, while remaining compatible with task accuracy measures. The dimensions we propose in our study provide concrete entry points for such user studies, where evaluation through a multidimensional assessment method for understanding would be both valid and worthwhile, with potential to inform other research contexts related to visualization and data understanding. 

Finally, MdamV could incorporate an educative feedback component. Respondents frequently expressed curiosity about their performance in the data reading tasks during qualitative testing. Providing immediate feedback, correct answers and explanations could enhance the assessment's value, encourage reflection, while serving literacy-building~\cite{chevalier2018}, and could function as an incentive for respondents.

\section{Conclusion}
With MdamV we contribute a method for assessing situated visualization understanding, allowing a range of abilities related to visual data literacy to be evaluated in direct relation to particular visualizations. By combining task-based measures, self-perceived ability ratings, and open-ended question formats, MdamV extends beyond the traditional focus on performance measures. Grounded in a theory-driven approach, MdamV comprises six distinct dimensions of visualization understanding, in reference to learning sciences frameworks. Validation was supported through a representative survey for Austria's population (ages 18-74, male/female split). This sample provided nuanced insights into the visualization understanding of a general audience, using two climate data visualizations, and revealed self-reported deficits such as perceived difficulties with simple measurement units and unfamiliarity with common chart types. Overall, the evaluation of MdamV suggests that its dimensions provide a valuable combination for assessing facets of visualization understanding. Importantly, MdamV points toward more nuanced and critical methodological approaches for assessing how general audiences interpret data visualizations. This, in turn, advances our understanding of how visualizations can be meaningfully designed and used in public contexts. 

\section*{Figure Credits}
\label{sec:figure_credits}
\begin{itemize}
    \item \autoref{fig:line-chart}: Statista, \textit{Der Meeresspiegel steigt kontinuierlich an}, NASA, https://de.statista.com/infografik/21922/anstieg-des-meeresspiegels/
    \item \autoref{fig:bar-chart}: Statista, \textit{Diese Länder stoßen am meisten $CO_2$ aus}, Global Carbon Project, https://de.statista.com/infografik/18287/co2-emissionen-in-ausgewaehlten-laendern/
\end{itemize}





\bibliographystyle{abbrv-doi-hyperref}

\bibliography{bib.bib}
\end{document}